\documentclass[conference]{IEEEtran}

\ifCLASSINFOpdf
\else
\fi

\usepackage{graphicx}
\usepackage[cmex10]{amsmath}
\usepackage{fixltx2e}
\usepackage{booktabs}
\usepackage{glossaries}
\usepackage[ruled]{algorithm2e} 
\usepackage{xcolor}
\usepackage[export]{adjustbox}
\usepackage{mathtools}
\usepackage{cite}
\usepackage{placeins}
\usepackage{stfloats}

\begin{document}
\title{ Interference-Aware Decoupled Cell Association in Device-to-Device based 5G Networks }

\author{\IEEEauthorblockN{Hisham Elshaer\IEEEauthorrefmark{1}\IEEEauthorrefmark{3}, Christoforos Vlachos\IEEEauthorrefmark{3}, 
Vasilis Friderikos\IEEEauthorrefmark{3} and Mischa Dohler\IEEEauthorrefmark{3}}
\IEEEauthorblockA{\IEEEauthorrefmark{1}Vodafone Group R\&D, Newbury, UK. }
\IEEEauthorblockA{\IEEEauthorrefmark{3}Centre for Telecommunications Research, King's College London, UK.\\
\textit{Email}: hisham.elshaer@vodafone.com \\
\{christoforos.vlachos, vasilis.friderikos, mischa.dohler\}@kcl.ac.uk}

}

\maketitle

\begin{abstract}
Cell association in cellular networks is an important aspect that impacts network capacity and eventually quality of experience. The scope of this work is to investigate the different and generalized cell association (CAS) strategies for Device-to-Device (D2D) communications in a cellular network infrastructure. To realize this,  we optimize D2D-based cell association by using the notion of uplink and downlink decoupling that was proven to offer significant performance gains. We propose an integer linear programming (ILP) optimization framework to achieve efficient D2D cell association that minimizes the interference caused by D2D devices onto cellular communications in the uplink as well as improve the D2D resource utilization efficiency. Simulation results based on Vodafone's LTE field trial network in a dense urban scenario highlight the performance gains and render this proposal a candidate design approach for future 5G networks.  
\end{abstract}

\begin{IEEEkeywords}
Cell association, decoupling, Device-to-Device, optimization, interference, resource utilization.
\end{IEEEkeywords}

\IEEEpeerreviewmaketitle

\section{Introduction}

The ever increasing cellular network traffic has led to a shift from single-tier homogeneous networks to multi-tier heterogeneous networks (HetNets) in an attempt to increase the network capacity in hotspots in an efficient and scalable way. The HetNet solution helps in improving the capacity of cellular networks and bringing the network closer to the user equipments (UEs). Device-to-Device (D2D) communication introduces similar benefits arising from the proximity of UEs to each others that is exploited by enabling direct communication between UEs without the need for the data to be routed via the fixed infrastructure network \cite{Lin2014}.

Until the fourth generation of cellular networks, cell association has been based on the downlink (DL) received signal power only. It was shown in \cite{Hisham2014} that associating both uplink (UL) and DL based on the DL received power in a HetNet is highly suboptimal and that the decoupling of both UL and DL results in substantial gains in the UL. D2D UEs are expected to have cellular and direct D2D communications in subsequent time instants or subframes. Therefore, D2D cell association needs to take the nature of cellular transmission into account. As per 3GPP \cite{3gpp.36.843}, D2D communication will take place in the UL licensed band which makes the decoupled association strategy very relevant to the D2D cell association problem. To the best of our knowledge, cell association has been extensively studied in macro-cellular systems, but only recently in heterogeneous networks \cite{Liu2014}. However, D2D-aware cell association is still an open issue for research and needs to be well investigated \cite{Vlachos2015}.     

The aim of this proposal is to study the different cell association (CAS) strategies for inband D2D communications  in a heterogeneous network. D2D technology is expected to yield numerous overall benefits that mainly arise from the proximity gain they offer. Therefore, meticulous enhancements need to be included that will make full use of its merits. We focus on the inband overlay communication where D2D and cellular communications take part in the licensed frequency band and there is no overlap in resource block (RB) usage between D2D and cellular communication. The contribution of this work is the optimization of D2D-based cell association using the decoupled UL and DL association concept developed in prior art\cite{Hisham2014}. To this end, integer linear programming (ILP) optimization tools are applied to introduce efficient D2D cell association that aims at minimizing the interference caused by D2D devices onto cellular communications as well as improve the efficiency of D2D resources usage.    

Considering the D2D communication paradigm, the interference management among D2D and cellular transmissions in inband is very challenging. Furthermore, commonly applied power control and interference management solutions within the literature usually resort to high complexity resource allocation methods, as stated in \cite{Asadi2014}. D2D and decoupled uplink and downlink have been identified as main building blocks of future 5G networks in \cite{BocHea14}. To this end, effective interference limitation with respect to resource utilization needs to be devised in order to improve the overall network welfare.

\begin{figure*}
  \includegraphics[width=\textwidth,trim = 0 1cm 0 1cm, clip = true]{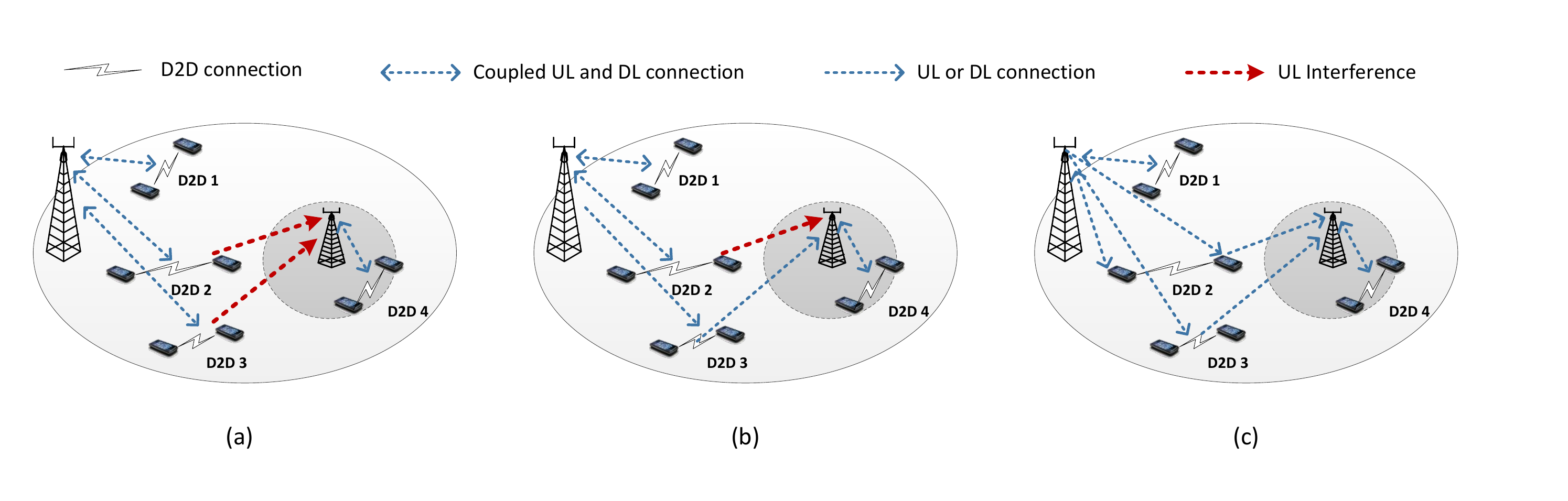}
  \caption{Considered cell association scenarios: (a) Joint-Coupled, (b) Joint-Decoupled, (c) Disjoint-Decoupled. }
\label{CA_scenarios}
\end{figure*}

\section{Problem Description}
\label{Prob_form}
The principle aim of this section is to firstly give a glimpse of the investigated cell association techniques and then dive into detailed analysis of their realization.  
Abiding by the milestones of LTE Release 12 and the prospective integration of D2D notion as a technological component to current and emerging networks, we present a number of general design assumptions according to up-to-date standardization working documents \cite{3gpp.36.843}:
\begin{itemize}
\item D2D connections will utilize the uplink (UL) resources.
\item The inband scenario of D2D communications is taken into consideration, where the interference from D2D devices onto cellular links in a neighbouring cell (either Macro-eNB (MeNB) or Small-eNB (SeNB)) could be substantial.
\item The transmit power of D2D devices will be controlled by the serving cell (MeNB or SeNB) based on fractional path-loss compensation power control \cite{3gpp.36.213} as done with cellular users.
Formally speaking, the transmit power of a D2D device $u$ associated with BS $b$ is given by
\begin{equation}
\label{eq:tx_power}
P_t^{lu} = min\{P_{Max}, 10\log_{10}(M) + P_0 + \alpha L_{blu}\},
\end{equation}
where $P_{Max}$ is the maximum transmit power of the device, $M$ is the number of physical resource blocks (PRB) assigned to the device, $P_0$ is a normalized power value (in dB), $\alpha$ is the pathloss compensation factor and $L_{blu}$ is the pathloss between the device $u$ of link $l$ and its serving cell $b$.
\end{itemize}

As mentioned in the previous section, D2D-based cell association algorithms need to consider the fact that D2D devices can have subsequent cellular and direct D2D transmissions in adjacent subframes to satisfy their communication needs. According to the current trend, a UE is primarily connected to a BS that provides  the highest DL received power. However, the decoupled UL and DL association proposed in \cite{Hisham2014} has shown substantial gains by allowing users to choose different cells in the UL and DL considering a cellular heterogeneous network. 

This idea constitutes the basis of this paper where it will be shown that the same concept is applicable to D2D-centric association as well. For the different cell association techniques that will be analysed, interference is the main validation criterion as it results from the ongoing cells' densification \cite{Ihalainen2013}.    

In the upcoming subsections, we provide an ILP optimization framework based on the different association policies by taking into account the notion of decoupling and the ability of the devices of a D2D pair to connect to different serving cells. Without loss of generality, unicast D2D connections are assumed. We will further compare these strategies in terms of transmit power efficiency, resource utilization and interference metrics. The considered cases are listed below.
\begin{itemize}
\item \textbf{Joint-Coupled (JC)}: The baseline case where devices of the same D2D pair are only allowed to connect to one cell (Joint). Furthermore, the D2D devices have the same UL and DL serving cell based on the DL received power (Coupled).
\item \textbf{Joint-Decoupled (JD)}: The devices of a D2D pair connect to the same serving cell but the UL and DL associations are decoupled where the UL serving cell is the one that minimizes the UL interference to cellular communication. 
\item \textbf{Disjoint-Decoupled (DD)}: The devices of a D2D pair are allowed to connect to different serving cells with the same association technique as the Joint-Decoupled case. 
\item \textbf{Hybrid-Decoupled (HD)}: In this case we combine both the Joint-Decoupled and the Disjoint-Decoupled cases to strike a balance between minimizing the interference and the resource usage. 
\end{itemize}

Considering the three last cases, we allow D2D UEs to be associated with different serving cells in the UL (decoupled access) based on the minimum UL interference metric. For the rest of the paper we will focus on the UL association optimization for the involved D2D UEs. 
To this end, before we detail the applied optimization framework, we need to define the set of deployed BSs as $B$ (including both MeNBs and SeNBs), the set of randomly distributed D2D links $\mathcal{L}$, and lastly, $U$ is the set of UEs that constitute these links.   

\subsection{Joint-Coupled CAS}
\label{suA}
In this scheme we assume that D2D UEs that constitute a link are associated with the same BS according to DL received power estimations. This approach is the baseline method as it is the technique used in LTE. However, the interference exerted by the D2D UEs that follow this association method can cause harmful effects on the cellular links, as clearly illustrated in Fig. \ref{CA_scenarios}a. In this figure, \textit{D2D 2} and \textit{D2D 3}, both associated (coupled) with the related MeNB can severely interfere with the proximate SeNB UEs active transmissions.

\subsection{Joint-Decoupled CAS}
\label{suB}
This scheme is realized by associating the D2D links in the UL with the BS minimizing the link's UL interference. Fig. \ref{CA_scenarios}b represents this case. In this scenario \textit{D2D 3} is served in the UL by the SeNB which results in the reduction of the transmit power of \textit{D2D 3} as the couple is closer to the SeNB. However due to the joint association constraint, \textit{D2D 2} is still associated to the MeNB.

Applied in the authors' prior work \cite{Vlachos2015}, we extend the cell association optimization logic for D2D links, where the paired devices are both connected to the same serving BS \cite{Yilmaz2014}. For this reason, we define the following binary decision variable
\begin{equation}
\label{eq:dec_var1}
y_{bl} = 
\begin{cases}
1, & \text{if D2D link } l \text{ is associated with BS } b \\ 
0, & \text{otherwise.}
\end{cases}
\end{equation}

Further, in order to view the problem of interference minimization caused by D2D UEs' potential transmissions, we need to define as $I_{bl} = mean\big\{I_{blu_1}, I_{blu_2} \big\}$ the average of the maximum interference generated by the two paired devices ($u_1$ and $u_2$ of link $l$) which are both associated with BS $b$. The corresponding interference term for a D2D device $u$ of link $l$ is given by $I_{blu} = max(P_t^{lu} \textbf{G}_{B'lu})$ where $\textbf{G}_{B'lu}$ is the matrix of link gains between user $u$ and all BSs  that belong to the set $B' = B -{b}$. Herein, $P_t^{lu}$ accounts for the transmission power of the UE $u$ of link $l$ according to \eqref{eq:tx_power} and depends on its associated BS.

The interference-based optimization problem can be then formulated as follows
\begin{equation}
\label{eq:joint_decoupled_opt}
\text{min} \:\:  \sum_{b \in B} \sum_{l \in \mathcal{L}} I_{bl} y_{bl}
\end{equation}
\begin{IEEEeqnarray}{ll}
\text{s.t.} ~
\sum_{b \in B} y_{bl} = 1, ~ \forall l \in \mathcal{L} \IEEEyessubnumber \label{eq:joint_decoupled_opt_con1}\\
\sum_{l \in \mathcal{L}} y_{bl} \leq K_b, ~ \forall b \in B \IEEEyessubnumber \label{eq:joint_decoupled_opt_con2}\\
y_{bl} \in \{0,1\}, ~ \forall b \in B, l \in \mathcal{L} \label{eq:joint_decoupled_opt_con3} \IEEEyessubnumber
\end{IEEEeqnarray} 
where constraint (\ref{eq:joint_decoupled_opt_con1}) requires the sole association of a D2D link $l$ to BS $b$, and (\ref{eq:joint_decoupled_opt_con2}) provides an upper bound of the number of user links that can be associated with every BS $b$. The difference of this scheme compared to the Joint-Coupled baseline strategy is the decoupling of DL and UL for the D2D links located in the topology. Intuitively, but as also proven in the sequel, this method is very efficient in terms of resource utilization by blocking (utilizing) one RB only from its associated BS that controls the D2D transmission. On the other hand, this method lacks intelligence in terms of interference controllability as it associates both devices of a D2D link to one BS without giving the flexibility for separate association of the nodes that could be less harmful.   

\subsection{ Disjoint-Decoupled CAS}
\label{suC}
In this decoupled D2D scenario the paired devices can be also connected to different serving cells as shown in Fig. \ref{CA_scenarios}c. We anticipate that, in terms of interference, this is a very efficient strategy as every device connects to its closest serving BS. However, this scheme is not efficient in terms of resource usage, simply because if the devices of a D2D pair are connected to two different BSs, the resources used by these devices have to be allocated (blocked) for the D2D connection in both cells as opposed to the case when both devices are served by the same BS where the resources will be allocated (blocked) only in one cell. Therefore this scheme is interference optimal but it uses twice as much resources as the Joint schemes. To this end, we provide an optimization setting that aims to minimize the introduced interference caused by the D2D transmissions. 

First, we consider the following binary decision variable that indicates each UE's association with a BS
\begin{equation}
\label{eq:dec_var1}
y_{blu} = 
\begin{cases}
1, & \text{if user } u \text{ of link } l \text{ associates with BS } b \\ 
0, & \text{otherwise.}
\end{cases}
\end{equation}
where  $b \in B$, $l \in \mathcal{L}$, and $u \in U$. 
Additionally, each link $l$ constitutes a direct link between two proximate devices (i.e. D2D devices $u_1$ and $u_2$) that, as already mentioned, can be either both associated with a serving BS \cite{Vlachos2015} or disjointly (loosely) connected with two separate BSs.

Therefore, the interference minimization problem for the disjoint decoupled D2D cell association can be set as follows 
\begin{equation}
\label{eq:dis_decoupled_opt}
\text{min} \:\:  \sum_{b \in B} \sum_{l \in \mathcal{L}} \sum_{u \in U} I_{blu} y_{blu}
\end{equation}
\begin{IEEEeqnarray}{ll}
\text{s.t.} ~
\sum_{b \in B} y_{blu} = 1, ~ \forall l \in \mathcal{L}, ~u \in U \IEEEyessubnumber \label{eq:dis_decoupled_opt_con1}\\
\sum_{l \in \mathcal{L}} \sum_{u \in U} y_{blu} \leq K_b, ~ \forall b \in B \IEEEyessubnumber \label{eq:dis_decoupled_opt_con2}\\
y_{blu} \in \{0,1\}, ~ \forall b \in B, l \in \mathcal{L}, u \in U \label{eq:dis_decoupled_opt_con3} \IEEEyessubnumber
\end{IEEEeqnarray} 
where $I_{blu}$ is the maximum interference generated by user device $u$ of D2D link $l$ if associated with BS $b$; its power part is again estimated according to \eqref{eq:tx_power}.

\subsection{Hybrid-Decoupled CAS}
\label{suD}
In this case, we propose an interference-aware optimization problem with an objective to achieve resource usage efficiency. An effective and controllable resource utilization on top of an interference-aware method may well entail in balanced interference mitigation and resource efficiency impact. The Disjoint-Decoupled approach might be optimal in terms of interference but it is not efficient in terms of resource usage. On the other hand, the Joint-Decoupled approach is optimal in the sense of resource usage but lacks of satisfactory interference performance compared to the two methods mentioned above. Hence, the Hybrid-Decoupled problem tries to strike the balance between interference and resource utilization.  
 
In order to realize this hybrid problem, an additional decision variable needs to be defined that will act as an indication of joint association for two devices that construct a D2D pair. This can be written as follows 
\begin{equation}
\label{eq:dec_var2}
z_{bl} = 
\begin{cases}
1, & \text{ if link } l \text{ associates with BS } b \\ 
0, & \text{otherwise.}
\end{cases}
\end{equation}
 Therefore, we propose a resource usage optimization problem that considers interference and formulate it as follows

\begin{equation}
\label{eq:hybrid_decoupled}
\text{max} \:\: \sum_{b \in B} \sum_{l \in \mathcal{L}} z_{bl}
\end{equation}
\begin{IEEEeqnarray}{ll}
\text{s.t.} ~ 
\sum_{b \in B} y_{blu} = 1, ~ \forall l \in \mathcal{L}, ~u \in U \IEEEyessubnumber \label{eq:hybrid_decoupled_con1}\\
\sum_{l \in \mathcal{L}} \sum_{u \in U} y_{blu} \leq K_b, ~ \forall b \in B \IEEEyessubnumber \label{eq:hybrid_decoupled_con2}\\
\sum_{u \in U} I_{blu} z_{bl} \leq I_{th},  ~ \forall b \in B, l \in \mathcal{L} \IEEEyessubnumber \label{eq:hybrid_decoupled_con3}\\
2x_{bl} \leq \sum_{u} y_{blu},  ~ \forall b \in B, l \in \mathcal{L} \IEEEyessubnumber \label{eq:hybrid_decoupled_con4}\\
\sum_{b \in B} z_{bl} \leq 1,  ~ \forall l \in \mathcal{L} \IEEEyessubnumber \label{eq:hybrid_decoupled_con5}\\
y_{blu}, z_{bl} \in \{0,1\}, ~ \forall b \in B, l \in \mathcal{L}, u \in U \IEEEyessubnumber \label{eq:hybrid_decoupled_con6}
\end{IEEEeqnarray} 
As shown, the main objective is the maximization of the number of joint connections for the distributed D2D paired devices with respect to interference. Constraints \eqref{eq:hybrid_decoupled_con1} and \eqref{eq:hybrid_decoupled_con2} are defined as in problem \eqref{eq:dis_decoupled_opt}. In \eqref{eq:hybrid_decoupled_con3}, a threshold that constrains the levels of interference if the devices of a link are jointly connected to a BS is added. This threshold can act as a weighting factor to decide if the focus of the algorithm should be interference (low $I_{th}$) or resource efficiency (high $I_{th}$). For this constraint, we limit the search to the $n$ closest BSs to reduce the search space and consequently the complexity and size of the inequality matrix. Furthermore, constraint \eqref{eq:hybrid_decoupled_con4} indicates that only if both devices of a link $l$ will be associated with the same BS $b$, the value of $z_{bl}$ variable equals to one (joint case). Lastly, \eqref{eq:hybrid_decoupled_con5} stands for the restriction that each link's users can be associated with only one BS in the case of joint connection ($z_{bl} = 1$). Differently, they are disjointly connected to two separate BSs ($z_{bl} = 0$).


\section{Simulation setup}   
\begin{figure}
\centering
\includegraphics[width = 8cm]{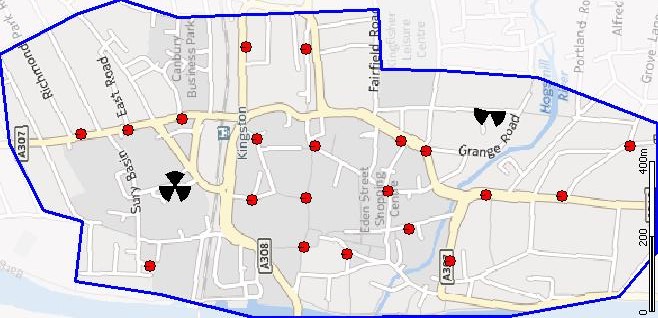}
\caption{Vodafone Small cell LTE test network.}
\label{topology}
\end{figure}

As deployment setup, we use the Vodafone LTE small cell test bed network deployment shown in Fig. \ref{topology}. The test network covers an area of approximately one square kilometre and includes two Macro sites and 21 SeNBs represented by the black shapes and red dots respectively. We use this existing test bed to simulate a relatively dense HetNet scenario. The propagation model is based on a high resolution 3D ray tracing pathloss prediction model. This model takes into account clutter, terrain and building data and it guarantees a realistic and accurate propagation model. The user distribution is based on real traffic data extracted from the live network. 
We assume an inband operation of D2D where D2D UEs use the same UL frequency band assigned for cellular transmission. However, D2D and cellular UEs are scheduled on different resources which is termed as 'overlay' operation in the literature.
The results are based on Monte Carlo simulations where the results are averaged over 100 simulation runs.

The operating frequency is 2.6 GHz. The maximum transmit powers of Mcells, Scells and UEs are 46, 30 and 23 dBm respectively. The fractional pathloss compensation power control algorithm in (\ref{eq:tx_power}) is assumed with $P_0 = -90 $ dBm and $\alpha = 0.8$. An average number of links of 336 is considered. $I_{th}$ is set to -130 dB. The next section features a set of results evaluating the proposed cell association methodologies proposed in Section \ref{Prob_form}. Finally, we assume that each D2D pair is allocated one resource block (RB) per base station.

\section{Simulation results}
\begin{figure}
\centering
\includegraphics[width=1\columnwidth]{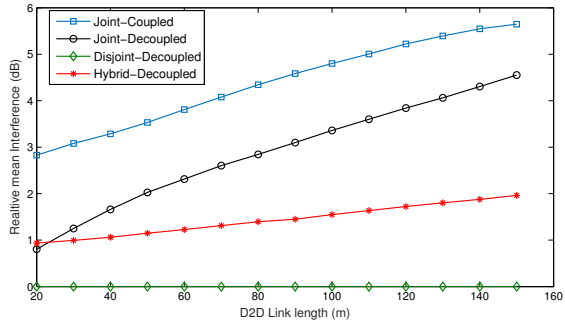}
\caption{Mean UL interference from D2D devices onto cellular transmissions.}
\label{mean_interf}
\end{figure}

In this section, a set of numerical evaluations is presented to investigate the proposed optimization schemes.  Fig. \ref{mean_interf} shows the mean UL interference exerted by the D2D UEs onto cellular UL transmission against the D2D link length. The interference  values are normalized relative to the DD case to show the different interference levels compared to this interference optimal scheme. The JC and JD schemes show an increasing interference trend with the link length where the interference levels are around 3 dB (twice) and almost 6 dB (4 times) more than the DD scheme at 100 m and 150 m link length respectively. This is logical as the more the link length increases the more suboptimal the joint association schemes are as forcing distant devices to connect to the same BS results in a higher transmit power of these devices and a higher interference to neighbouring cells. The HD scheme introduces a trade-off between the Joint and DD schemes as it maintains an almost constant interference level that is around 1 dB higher than the DD scheme. This is due to the intelligence in the HD scheme that allows it to jointly/dis-jointly allocate D2D pairs depending on the interference level.

\begin{figure}
\centering
\includegraphics[width=1\columnwidth]{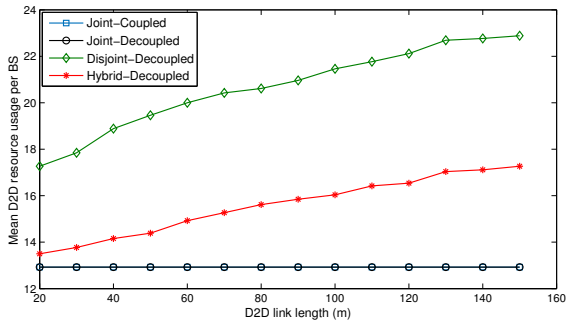}
\caption{Mean resource utilization for D2D per base station.}
\label{resource_u}
\end{figure}

As explained earlier, if a D2D pair is served by one BS it is assumed to use only one RB over the whole network as this resource is reserved for this D2D pair in this BS only. However, if the devices of a pair are associated to different BSs then it is assumed that this pair is using two RBs over all the network since one RB has to be allocated for that pair in both BSs.
Fig. \ref{resource_u} illustrates the average D2D resource usage per BS against the link length. The figure shows a constant resource usage for the JC and JD association schemes. This trend can be explained by the fact that the D2D pairs are jointly associated to the same BS regardless of the link length. Hence each D2D link uses 1 RB independent of the link length. However, the DD scheme shows an increasing RB usage with the link length.  This can be explained by the fact that the probability of disjoint association increases with the link length and so as the D2D resource usage in the whole network since the disjoint D2D link uses twice as much RBs as the joint one. The HD scheme -again- offers a compromise between the joint and disjoint schemes as the main scope of the optimization problem is to improve the resource usage efficiency with a constraint on the interference. The HD method achieves a reduction of resource usage of about 45\% at 150 m link length compared to DD.
Thus, it can be noted that the HD scheme offers a trade-off between the UL interference and resource efficiency which can be controlled by setting the $I_{th}$ accordingly.
 
\begin{figure}
\centering
\includegraphics[width=1\columnwidth]{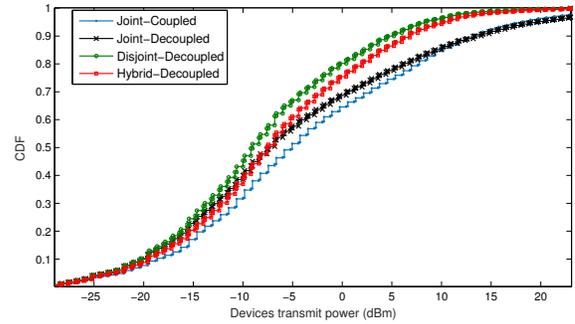}
\caption{CDF of the devices' transmit power.}
\label{Power_CDF}
\end{figure}

The cumulative distribution function (CDF) of the D2D UEs transmit power is shown in Fig. \ref{Power_CDF}. The figure shows that the JC and DD schemes have the highest and lowest transmit power distributions with a difference of more than 5 dB at 50\% of the CDF which increases the higher the transmit power is. The HD has a distribution that fits mid-way between the JC and JD distribution and that gets closer to the DD the higher the transmit power is. This shows that the HD scheme can result in a reduction of transmit power that varies between 3-5 dBs which is deemed crucial for battery powered devices.

\section{Conclusions}
In this paper, we presented different cell association strategies for D2D communications in cellular networks. Based on the notion of decoupled UL and DL connections, we proposed an integer linear programming (ILP) optimization framework that aimed at achieving efficient D2D cell association with respect to interference reduction as well as resource utilization. Extensive simulations show the significant gains of the applied methods. The Disjoint-Decoupled (DD) optimization technique achieves more than twice in UL interference reduction as well as more than 5 dBs reduction in devices transmit power compared to baseline cell association methods. However, DD results in an inefficient use of D2D resources.  Therefore, we introduced the Hybrid-Decoupled (HD) technique which achieves a balance between the interference reduction and resource utilization efficiency. HD results in a slightly worse interference performance than the DD scheme but with a 45\% improvement in resource usage efficiency. We deem HD to be a strong candidate for future 5G cell association algorithms.

\section*{Acknowledgment}

This research is co-funded by Vodafone Group R\&D and CROSSFIRE MITN Marie Curie project (FP7-PEOPLE-317126).

\bibliographystyle{IEEEtran}
\bibliography{decoupledD2D_references}

\end{document}